\documentclass[%
preprint, 
 amsmath,amssymb,
 aps, 
 pra,
 showkeys, 
]{revtex4-2}

\usepackage[version=4]{mhchem} 
\usepackage{makecell}          
\usepackage{amsmath}           
\usepackage{graphicx}          
\usepackage{dcolumn}
\usepackage{bm}
\usepackage{xcolor}

\begin{document}

\preprint{APS/123-QED}

\title{\textbf{Effects of gravity on lean hydrogen/air flame instability: From linear scaling law to nonlinear morphology evolution}}

\author{Qizhe Wen}
\affiliation{SKLTCS, HEDPS, School of Mechanics and Engineering Science, Peking University, Beijing, 100871, P.R. China}

\author{Yan Wang}
\affiliation{SKLTCS, HEDPS, School of Mechanics and Engineering Science, Peking University, Beijing, 100871, P.R. China}

\author{Linlin Yang}
\affiliation{SKLTCS, HEDPS, School of Mechanics and Engineering Science, Peking University, Beijing, 100871, P.R. China}

\author{Yiqing Wang}
\affiliation{Department of Aeronautical and Aviation Engineering, The Hong Kong Polytechnic University, Hong Kong, 999077,  P.R. China}

\author{Thorsten Zirwes}
\affiliation{Institute for Reactive Flows (IRST), University of Stuttgart, Pfaffenwaldring 31, Stuttgart, 70569, Germany}

\author{Shengkai Wang}
\email{Corresponding author: sk.wang@pku.edu.cn (Shengkai Wang)}
\affiliation{SKLTCS, HEDPS, School of Mechanics and Engineering Science, Peking University, Beijing, 100871, P.R. China}

\author{Zheng Chen}
\affiliation{SKLTCS, HEDPS, School of Mechanics and Engineering Science, Peking University, Beijing, 100871, P.R. China}

\date{\today}

\begin{abstract}
The instability characteristics of lean hydrogen/air flames have attracted considerable research attention, yet the effect of gravity remains insufficiently understood. In this study, time-resolved two-dimensional simulations with detailed chemistry and transport are conducted to investigate the influence of gravity-induced Rayleigh-Taylor (RT) instability on the linear growth rate of disturbances and nonlinear morphology evolution of cellular flame fronts at different length scales. In the linear regime, a parametric study is performed across various equivalence ratios, initial temperatures and pressures; in each case, the dispersion relation is calculated for various gravity levels. The influence of gravity is most pronounced under ultra-lean, low-temperature, and high-pressure conditions, and a universal scaling law between gravity sensitivity and the Froude number is established. In the nonlinear regime, gravity has opposite effects on the large-scale and small-scale structures of lean hydrogen flames. On the one hand, positive gravity inhibits the splitting of small-scale cellular structures through a baroclinic torque mechanism; on the other hand, it promotes the development of large-scale finger-like structures, thereby increasing the total surface area and the global consumption speed of the flame. The effects of gravity on the probability distributions of cell size, displacement speed, Karlovitz number, and local curvature are also analyzed. The results and findings of the present study should advance the fundamental understanding of hydrogen flame dynamics under varying gravity conditions and provide insight for relevant applications, including fire safety and space propulsion.
\end{abstract}

\maketitle
\clearpage

\section{Introduction\label{sec:introduction}} \addvspace{10pt}
Hydrogen has drawn increasing attention as a zero-carbon alternative to fossil fuels ~\cite{mallapaty2020china}. Lean hydrogen/air flames are particularly advantageous due to their lower flame temperatures and reduced \(\text{NO}_\text{x}\) emissions. However, hydrogen's large mass diffusivity and high thermal expansion ratio give rise to multiple intrinsic instability mechanisms. A comprehensive understanding of these mechanisms is crucial for the development of advanced hydrogen combustion systems.

There are three primary instability mechanisms in hydrogen/air flames: hydrodynamic, thermal-diffusive (TD), and Rayleigh–Taylor (RT) instabilities. Hydrodynamic instability, commonly referred to as Darrieus–Landau (DL) instability \cite{ darrieus1938propagation,landau1944theory}, arises from the density jump across the flame front and is particularly pronounced under cryogenic conditions~\cite{yang2025non}. TD instability results from the coupling between flame stretch and preferential diffusion of heat and mass (also known as the Lewis number effect) ~\cite{barenblatt1962heat}, which plays a critical role in lean flames of hydrogen. Over the past several decades, DL and TD instabilities, along with their coupling effects, have been extensively investigated through theoretical analyses~\cite{sivashinsky1977diffusional, clavin1983influence, matalon2003hydrodynamic, law2010combustion}, experimental measurements ~\cite{LAW2005Cellular, yan2026quantitative, clanet1998first}, and numerical simulations ~\cite{altantzis2012hydrodynamic, berger2022intrinsic_a, berger2023flame, lapenna2024synergistic, gaucherand2023intrinsic}, leading to a comprehensive understanding of both disturbance growth rates in the linear regime and the evolution of complex cellular structures in the nonlinear regime.

Compared with DL and TD instabilities, considerably less attention has been given to the quantitative influence of the RT instability, which is driven by a gravitational or other body-force field. Theoretical efforts have sought to establish analytical dispersion relations to describe the linear growth rate of the overall instability arising from all three mechanisms. Notably, Pelce and Clavin \cite{pelce1982influence} demonstrated that downward-propagating flames can fully suppress inherent hydrodynamic wrinkling only when the Froude number falls below a critical threshold. Furthermore, comprehensive dispersion relations including gravity effects were developed by Clavin and Garcia \cite{clavin1983influence} and extended by Law \cite{law2010combustion}. However, although these analytical frameworks provide important physical insights, their derivations inherently rely on the assumption of a near-unity effective Lewis number to achieve mathematical closure, which fundamentally limits their applicability for actual lean hydrogen mixtures.

Experimentally, previous studies under both normal gravity (1-g) and microgravity ($\mu$-g) conditions have yielded invaluable insights, despite the challenges in isolating the gravity-dependent RT mechanism from other coupled instabilities. Early microgravity tests in drop towers and parabolic flights (e.g., \cite{ronney1990near, dunsky1992microgravity}) have confirmed that, in the absence of gravity, premixed flames of effective Lewis numbers less than unity are subject to inherent hydrodynamic wrinkling and can develop highly curved cellular structures, while in downward propagating flames under normal gravity such wrinkling is suppressed. More recently, experiments in Hele-Shaw cells (e.g., \cite{sarraf2018quantitative, battikh2023nonlinear}) have been utilized as an alternative approach to reduce the effect of buoyancy on flame propagation. These experiments significantly extended the test time and enabled direct observation of gravity effects in the highly nonlinear regime, for example, the modification of flame brush amplitude and cell merging dynamics by buoyancy. However, since these experiments were subject to heat loss and friction near the walls, isolating the gravity influence from other effects remained challenging. 

In light of these challenges, numerical simulations have been extensively employed as a complementary approach to investigate the effect of gravity on flame instability. Early studies using two-dimensional simulations (e.g., \cite{oran1989time, patnaik1991effect_a, patnaik1991effect_b}) have verified the stabilizing/destabilizing effect of gravity under different flame propagation directions. In addition, the unstable wavenumber range of a flame was found to narrow during downward propagation and broaden during upward propagation \cite{kadowaki2001body}. Recently, the rapid advancement of computing power has enabled high-fidelity direct numerical simulations (DNS) in the highly nonlinear regime. For example, Wang et al. \cite{wang2024gravitational} conducted a DNS study of the large-scale cellular flames and revealed that gravity flattens the flame front and suppresses cusp fusion while promoting flame front splitting, leading to a nearly linear scaling of the global propagation velocity with the inverse Froude number. Fernandez et al. \cite{fernandez2019impact} numerically explored the narrow-channel configuration (similar to the Hele-Shaw cells) to isolate buoyancy effects. Tavares and Jayachandran \cite{tavares2024dynamics} quantified the RT-induced flame acceleration in slowly propagating refrigerant flames. In particular, for lean hydrogen/air mixtures, Zheng et al. \cite{zheng2024stability} investigated 2D laminar flames at $\phi$ = 0.3 and $\phi$ = 0.4 and found that the most-probable cell sizes are larger for RT-unstable flames as compared to RT-stable flames. Most recently, their study was extended to 3D turbulent regimes and multi-component fuel blends \cite{zheng2025rayleigh, zheng2026characteristics,ZHENG2026effects}, with a focus on RT-induced macroscopic turbulence characteristics, enstrophy transport, and NO$_x$ formation.

Despite these advances, several knowledge gaps remain at different stages of flame evolution under finite gravity. In the initial stage of small-amplitude disturbance evolution (the linear regime), accurately predicting the growth rate with analytical models is challenging due to simplifications in the underlying theories, and high-fidelity DNS calculations are computationally expensive and scarce. Consequently, the exact magnitude of gravity effects on flame instability has not been quantitatively assessed. In the later stage of fully developed instability (the nonlinear regime), where the flame front is populated with small-scale cellular and large-scale finger-like structures, the quantitative influence of gravity on flame morphology evolution across spatial scales, together with the underlying physical mechanisms, remains to be further elucidated.

The present study is motivated to quantify the effect of gravity on lean hydrogen/air flames in both the linear and nonlinear regimes. Specifically, the influence of gravity on the perturbation growth rate in the linear regime is discussed in Section~\ref{sec:linear}, and the influence of gravity on flame morphology evolution in the nonlinear regime and its scale-dependence are examined in Section ~\ref{sec:nonlinear}.

\section{Numerical methods} \addvspace{10pt}

\begin{figure}
    \centering
    \includegraphics[width=0.5\linewidth]{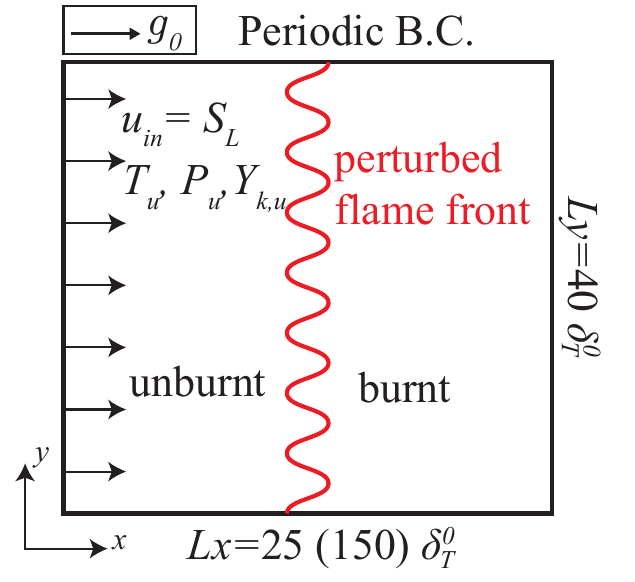}
    \caption{Computational setup and boundary conditions. $L_x$ and $L_y$ denote the domain lengths used for the linear (nonlinear) regime analyses. The flame front is illustrated with a single harmonic perturbation.}
    \label{fig:domain}
\end{figure}

In the present study, direct numerical simulations (DNS) of premixed hydrogen-air flames are conducted in a two-dimensional rectangular computation domain, following a numerical setup similar to that of Berger et al.~\cite{berger2019characteristic} and Zirwes et al.~\cite{zirwes2024role}. Both the linear and non-linear regimes are investigated. The schematic of the computation domain and boundary conditions is shown in Fig.~\ref{fig:domain}.
\begin{table*}[bht]
    \footnotesize
    \centering
    \caption{Four sets of simulation conditions used for the linear-regime analysis.}
    \begin{tabular}{lcccc}
        \hline
        Type & \(\phi\) [-] & \(T_u\)~[K] & \(P_u\)~[atm] & \(Fr =   S_L^2/(10g_0\delta_T^0)\) [-] \\ 
        \hline
        R0 - Reference& 0.4 & 300 & 1 & 0.86 \\ 
        R1 - Varying \(\phi\) & 0.3, 0.5, 1.0, 1.5 & 300 & 1 & 0.007, 6.46, 123.30, 240.26 \\ 
        R2 - Varying \(T_u\) & 0.4 & 200, 500, 700 & 1 & 0.08, 26.72, 319.53\\ 
        R3 - Varying \(P_u\) & 0.4 & 300 & 2, 3, 5 & 0.66, 0.45, 0.14 \\ 
        \hline
    \end{tabular}
    \label{tab:caselist_linear_regime}
\end{table*}

For simulations in the linear regime, the streamwise and lateral lengths of the computation domain are set as $L_x=25\delta_T^0$ and $L_y=40\delta_T^0$, respectively, where $\delta_T^0=(T_b-T_u)/\mathrm{max}(\left|\nabla T\right|)$ denotes the thermal thickness of a zero-stretch planar flame. To investigate the effect of Rayleigh-Taylor (RT) instability on the linear growth rate, four sets of simulation have been designed and are labeled R0 to R3 in Table \ref{tab:caselist_linear_regime}. R0 is the reference set with an equivalence ratio of $\phi=0.4$ and an initial temperature of $T_u=300\text{K}$. In set R1, $\phi$ is varied between 0.3 and 1.5 to investigate the influence of gravity on flame instability under lean, stoichiometric and rich conditions. In set R2, $T_u$ is varied over 200 - 700 K. In set R3, the ambient pressure $P_u$ is varied from 2 to 5 atm to assess the influence of elevated pressure on flame instability. For all simulation conditions, three gravity levels are considered: $g=0, +10g_0$, and $-10g_0$, where $g_0=9.8\text{ m/s}^2$ denotes standard gravity. Specifically, positive gravity ($+10g_0$) is antiparallel to the direction of flame propagation, representing an RT-unstable configuration, while negative gravity ($-10g_0$) corresponds to the RT-stable case. The corresponding Froude numbers \(Fr=S_L^2/(10g_0\delta_T^0)\) are also listed in Table \ref{tab:caselist_linear_regime}. The multi-wavenumber perturbation method ~\cite{al2024efficient} is employed to seed the cellular instability, enabling efficient and accurate calculation of the dispersion relation across wavenumbers. A small perturbation amplitude of $A_n=0.01\delta_T^0$ is used to ensure a sufficiently long linear regime. A uniform grid with spatial resolution $\Delta x = \Delta y = \delta^0_T / 20$ is employed to accurately resolve the small initial perturbations~\cite{zirwes2024role}. For cases with relatively large flame thickness, the grid is further refined so that the spatial resolution does not exceed $35\,\mu\text{m}$.

Simulations in the nonlinear regime focus on lean flames at $\phi$ = 0.4. A larger computation domain of $L_x=150\delta_T^0$ and $L_y=40\delta_T^0$ is used to capture the long-term evolution of the perturbed flame. The initial field is perturbed by a single-wavenumber disturbance with a wavelength of $\lambda=8\delta_T^0$~\cite{zirwes2024role} and an initial disturbance amplitude of $A_0=0.05\delta_T^0$. To simulate nonlinear flame evolution at long times, a coarser grid size of $\Delta x = \Delta y = \delta_T^0 / 10$ is adopted to improve the computational efficiency~\cite{berger2019characteristic, zheng2024stability, wen2023numerical}.

For all flame cases explored in the present study, simulations are performed using the EBIdnsFOAM solver developed by Zirwes et al.~\cite{zirwes2023assessment}. EBIdnsFoam has been extensively validated and used in previous studies; the corresponding numerical methods and code validations are presented in \cite{zirwes2023assessment} and are not repeated here. The simulations employ a mixture-averaged transport model that includes the Soret effect. In all cases, the hydrogen reaction mechanism developed by Li et al.~\cite{li2004updated} is used.

\section{Results and discussion} \addvspace{10pt}
\subsection{Gravity effects in the linear regime}\label{sec:linear}
\subsubsection{Dispersion relation}

\begin{figure*}[!htb]
\centering
\vspace{10 pt}
\includegraphics[width=\linewidth]{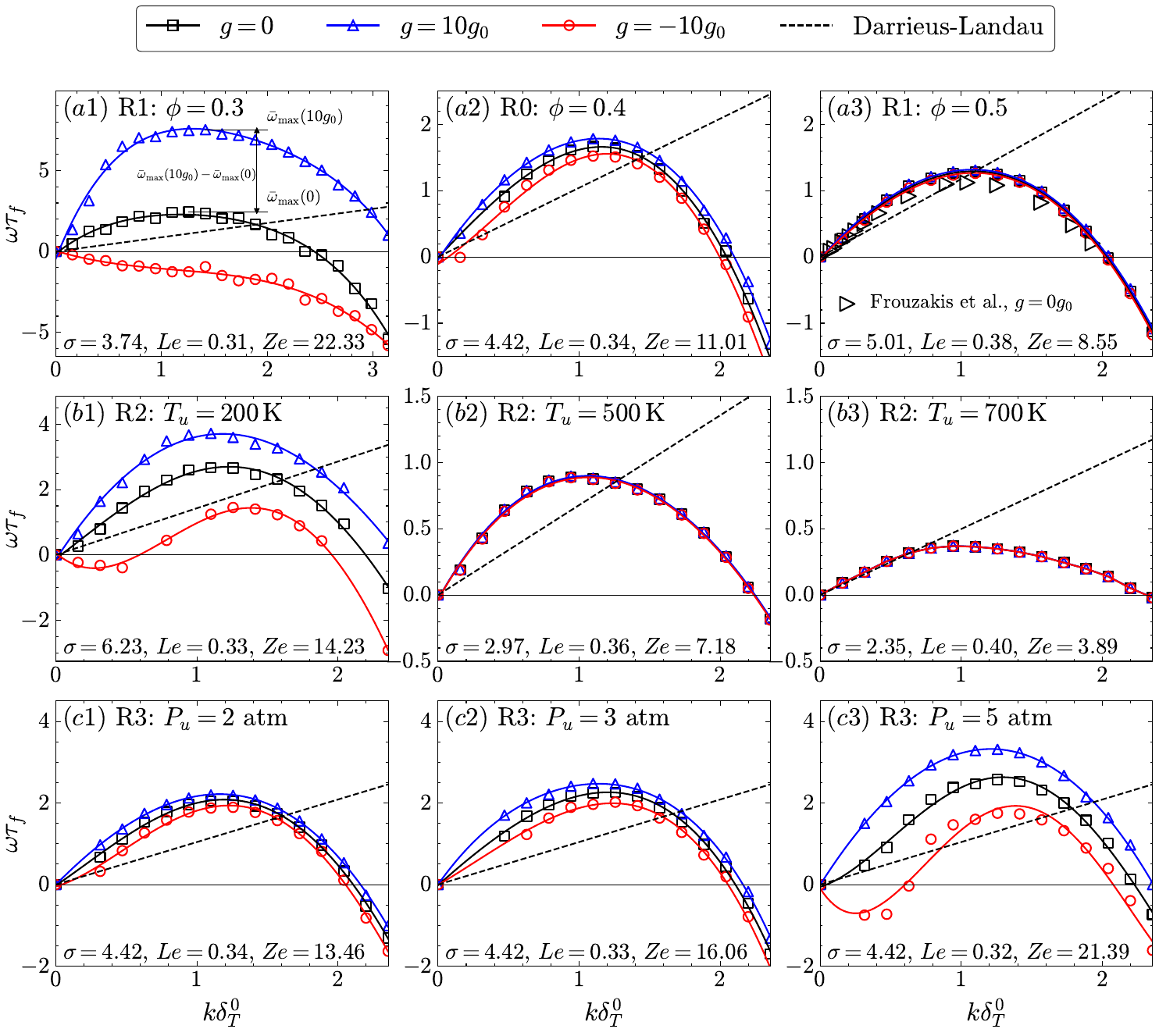}
\caption{Dispersion relations for Case R1 (varying \(\phi\), top row), Case R2 (varying \(T_u\), middle row) and Case R3 (varying \(P_u\), bottom row). The influence of gravity is more pronounced at lower equivalence ratios, lower unburned temperatures and higher pressures.}
\label{fig:six_fig}
\end{figure*}

Fig. \ref{fig:six_fig} illustrates the numerical dispersion relations for (a) various equivalence ratios (set R1), (b) unburned gas temperatures (set R2), and (c) pressures (set R3) under \(g=0\) and \(g = \pm 10g_0\). The DNS results are indicated by symbols, while the solid lines represent fourth-order polynomial fits to the data. The wavenumber and growth rate are nondimensionalized as $\bar{k} = k\delta_T^0$ and $\bar{\omega} = \omega\tau_f$, respectively, with $\tau_f = \delta_T^0 / S_L$ denoting the characteristic flame time. The accuracy of the current results is validated at a representative condition of $\phi$ = 0.5 by comparison with the previous work of Frouzakis et al. ~\cite{frouzakis2015numerical}, as shown in Fig. \ref{fig:six_fig}(a3), where excellent agreement is observed. 

The dispersion relation for the Darrieus–Landau (DL) instability is also computed in the present study using the following equation~\cite{darrieus1938propagation,landau1944theory}:
\begin{equation}
\bar{\omega}_{DL}(\bar k)=\frac{\sqrt{\sigma^3+\sigma^2-\sigma}-\sigma}{\sigma+1}\bar{k},
    \label{eqn:DL}
\end{equation}
where \( \sigma = \rho_u/\rho_b\) is the thermal expansion ratio. The results are also shown in Fig.~\ref{fig:six_fig}. For reference, the effective Lewis number $Le$ and the Zeldovich number $Ze$ are annotated in the figure. Physically, $Le$ represents the ratio of thermal diffusivity to mass diffusivity, and $Ze$ quantifies the sensitivity of chemical reaction rate to changes in flame temperature. Additional details on the numerical determination of $Le$ and $Ze$ are provided in the Supplementary Material.

It is worth noting that, although there have been various theoretical models for linear flame instability analysis, they generally rely on a one-step global reaction approximation that is not applicable to hydrogen flames. From a reaction-kinetics perspective, hydrogen flames are governed by two groups of reactions, chain-branching and chain-termination, which makes their reaction kinetics more complex than carbon-containing fuels (e.g., alkanes). This complexity is reflected in the distinctive Z-shaped explosion-limit curve of hydrogen compared to other fuels \cite{liu2023explosion}. Consequently, the present study relies on high-fidelity DNS with detailed chemistry to accurately determine the dispersion relations of lean hydrogen flames. A comparison between the present simulations and existing theoretical models is discussed in the Supplementary Material. 

Under zero gravity conditions, where the RT instability is absent (black curves), the dimensionless growth rate obtained from DNS exceeds the DL instability model at low wavenumbers. This is consistent with previous observations that the TD instability increases the growth rate at small wavenumbers ~\cite{berger2022intrinsic_a}. As shown in Fig. \ref{fig:six_fig}(a1)-(a3), the maximum growth rate $\bar \omega_{max}$ decreases as the equivalence ratio $\phi$ increases from ultra-lean ($\phi=0.3$) to lean ($\phi=0.5$) conditions. This trend is primarily due to the increase in the effective Lewis number, which suppresses preferential diffusion and thus weakens the TD instability. The initial temperature $T_u$ also affects the growth rate. As illustrated in Fig. \ref{fig:six_fig}(b1)-(b3), increasing $T_u$ reduces not only the thermal expansion ratio $\sigma$ (thereby weakening the DL instability) but also $Ze$, which further suppresses the TD instability \cite{berger2022intrinsic_a}. In contrast, increasing the ambient pressure promotes flame instability, as shown in Fig. \ref{fig:six_fig}(c1)-(c3). The pressure effect is twofold: (1) higher pressures decrease the flame thickness, which weakens the stabilizing effect of flame curvature~\cite{YUAN2007flame}; and (2) elevated pressure increases \(Ze\) and enhances the TD instability \cite{berger2022intrinsic_a}.

Results at finite gravity show that positive gravity amplifies the growth rate while negative gravity suppresses it, with the magnitude of the gravitational effect strongly depending on the operating conditions. For the ultra-lean mixture at $\phi=0.3$ (Fig. \ref{fig:six_fig}(a1)), the gravity effect is substantial: under negative gravity ($g=-10g_0$) the flame remains stable across all wavenumbers, whereas under positive gravity ($g=10g_0$), the maximum growth rate $\bar \omega_{max}$ more than doubles compared with the zero-gravity case. As $\phi$ increases, the relative influence of gravity diminishes; at $\phi=0.5$ (Fig. \ref{fig:six_fig}(a3)) the dispersion relation curves under different gravity levels nearly collapse onto a single curve, indicating a negligible gravity effect. 

A similar sensitivity is observed with respect to the initial temperature. For a cryogenic $T_u=200\text{ K}$ (Fig. \ref{fig:six_fig}(b1)), negative gravity ($g=-10g_0$) stabilizes the flame only in the low-wavenumber range ($\overline{k} < 0.5$),  consistent with the existing theories that buoyancy mainly suppresses large-scale perturbations \cite{law2010combustion}. In contrast, at $T_u=700\text{ K}$ (Fig. \ref{fig:six_fig}(b3)) gravity has little effect on the instability growth rate. Gravity sensitivity also increases with ambient pressure. For example, at $P_u=5\text{ atm}$ (Fig. \ref{fig:six_fig}(c3)) negative gravity is seen to stabilize perturbations at low wavenumbers.

\subsubsection{Universal scaling law for the gravity sensitivity of the maximum linear growth rate}

To quantify how gravity modulates flame instability across different unburned-gas conditions, we define a dimensionless global sensitivity parameter $\eta$ as follows:
\begin{equation}
    \eta = \frac{\bar{\omega}_\mathrm{max}(10g_0)-\bar{\omega}_\mathrm{max}(0)}{\bar{\omega}_\mathrm{max}(0)}
\end{equation}
In this equation, $\bar{\omega}_\mathrm{max}(10g_0)$ and $\bar{\omega}_\mathrm{max}(0)$ denote the maximum dimensionless growth rates at $g = 10g_0$ and $g = 0$, respectively, as illustrated in Fig.~\ref{fig:six_fig}(a1).

\begin{figure}[h]
\centering
\vspace{10 pt}
\includegraphics[width=0.5\linewidth]{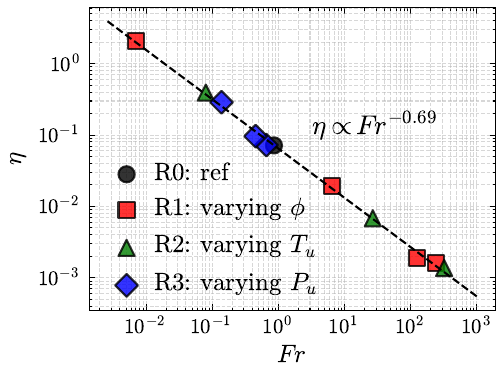}
\caption{The relationship between the global gravity-sensitivity parameter $\eta$ and the Froude number $Fr$ under various conditions. The solid line represents a power-law fit for all cases.}
\label{fig:Scaling_Law_Unified}
\end{figure}

For all cases explored in the current study, a unified correlation is observed between the global gravity-sensitivity parameter $\eta$ and the Froude number $Fr$ (defined at $g = 10g_0$), as shown in Fig. \ref{fig:Scaling_Law_Unified}. Specifically, $\eta$ scales approximately as $Fr^{-0.69}$. The negative dependence of $\eta$ on $Fr = S_L^2/g\delta_T^0 = \tau_b/\tau_f$ indicates that the gravity sensitivity of linear instability is governed by the relative time scale of buoyancy ($\tau_b = S_L/g$) to the flame ($\tau_f = \delta_T^0/S_L$). At $Fr > 10$, $\eta$ falls below 0.01, implying that the gravitational influence on flame stability becomes negligible. Additional analysis of the sensitivity at various gravity levels is provided in the Supplementary Material.

\subsection{Gravity effects in the nonlinear regime} \label{sec:nonlinear}

In this section, we investigate the effects of gravity on the dynamic evolution of cellular structures for flame cases in set R0~(\(\phi=0.4\)). A small sinusoidal perturbation is applied at time zero to trigger the exponential growth of flame instability. The long-time behavior of instability evolution in the nonlinear regime (for example, at $t>4\tau_f$) is relatively insensitive to the exact form of the initial perturbation, because it is dominated by interactions among the most energetic modes dictated by the dispersion relations. In the nonlinear regime, flame cells develop secondary structures that subsequently undergo splitting and merging, ultimately evolving into highly complex and chaotic flame morphologies \cite{zirwes2024role, Sivashinsky1977nonlinear, berger2022intrinsic_b, zheng2024stability}.

\subsubsection{Splitting of cellular flame structures} \label{sec:cell_split}
\begin{figure}[h!]
\centering
\vspace{10 pt}
\includegraphics[width=0.75\columnwidth]{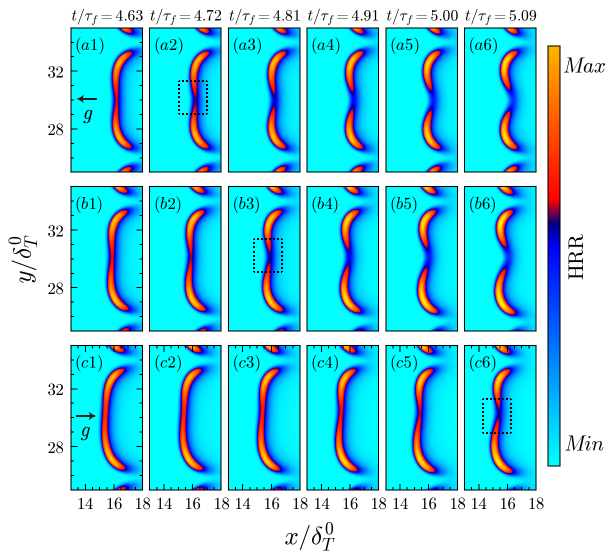}
\caption{Snapshots of the dynamic evolution and splitting of cellular structures for a flame at $\phi=0.4$. The temporal evolution highlights the cell breakup process under different gravity conditions: (a) $g=-10g_0$, with splitting at $t/\tau_f=4.72$; (b) $g=0g_0$, with splitting at $t/\tau_f=4.81$; and (c) $g=10g_0$, with splitting at $t/\tau_f=5.09$.}
\label{fig:cell_split}
\end{figure}

Fig. ~\ref{fig:cell_split} compares the initial cell-splitting dynamics under different gravity conditions. An interesting and somewhat counter-intuitive observation is the delay of cell splitting under positive gravity~(RT-unstable). Although one might expect gravity-induced RT instability to increase perturbation growth and fragmentation, the results show that positive gravity stabilizes the flame front and leads to more organized cellular patterns.

To elucidate this phenomenon, we perform additional analysis of the flame displacement speed $S_d$ and the flame stretch rate $K_s$. $S_d$ is defined as the velocity of the flame front relative to the local flow field. The flame front is identified by an isocontour corresponding to a specific progress variable $c=c^*$, where $c^*$ is the value at the peak heat-release rate for a planar flame at the same equivalence ratio. The progress variable $c$ is defined based on the \ce{H2} mass fraction as $c = (Y_{\ce{H2},u} - Y_{\ce{H2}}) / (Y_{\ce{H2},u} - Y_{\ce{H2},b})$, where the subscripts '$u$' and '$b$' denote the unburned and burned states, respectively. $S_d$ comprises three components corresponding to reaction, normal diffusion, and tangential diffusion, respectively:
\begin{equation}
    S_d = S_{d,r} + S_{d,n} + S_{d,t}
\end{equation}

Specifically, the reaction component, $S_{d,r} = \dot{\omega}_{c} / (\rho |\nabla c|)$, represents the chemical source term; the normal diffusion component, $S_{d,n} = \mathbf{n} \cdot \nabla (\rho D \mathbf{n} \cdot \nabla c) / (\rho |\nabla c|)$, describes diffusion normal to the flame front; and the tangential diffusion term, $S_{d,t} = -2D\kappa$, is explicitly coupled with the local flame curvature $\kappa = 0.5 \nabla \cdot \mathbf{n}$, where $\mathbf{n} = -\nabla c/|\nabla c|$ is the unit normal vector pointing toward the unburned mixture, and $D$ denotes the mass diffusivity.

Similarly, the flame stretch rate, defined as $K_s=(dA/dt)/A$, can be decomposed into two components:
\begin{equation}
    K_s = a_t + 2\kappa S_d,
    \label{eqn:Ks}
\end{equation}
where $a_t$ is the tangential strain rate, and $2\kappa S_d$ is the curvature stretch. The dimensionless stretch rate and its components are expressed via the Karlovitz numbers: $Ka_s=K_s\tau_f$, $Ka_{s,t}=a_t\tau_f$, and $Ka_{s,c}=2\kappa S_d\tau_f$.

\begin{figure}[h!]
\centering
\vspace{10 pt}
\includegraphics[width=0.75\columnwidth]{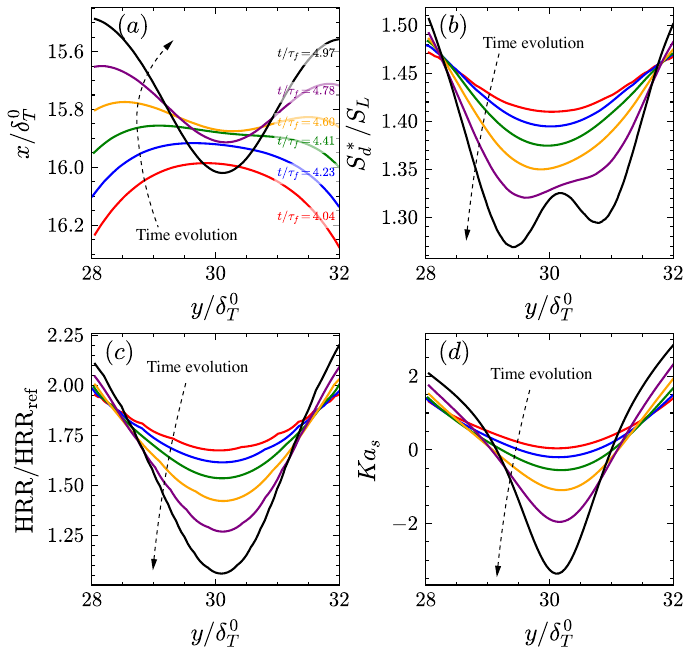}
\caption{Temporal evolution of single-cell splitting under zero-gravity ($g=0$) conditions. (a) flame front morphology, (b) displacement velocity \(S_d^*/S_L\), (c) heat release rate (HRR) normalized by the maximum HRR of a one-dimensional flame, and (d) Karlovitz number. Results at six selected times are shown in different colors, with red representing the earliest time and black the latest. The vertical axis in (a) is reversed for improved presentation.}
\label{fig:Sd-Y}
\end{figure}

Fig.~\ref{fig:Sd-Y} shows the temporal evolution of key scalar distributions during the early stage of a single-cell splitting process under zero-gravity conditions ($g=0$). Parameters displayed include the flame front location, density-weighted displacement speed $S_d^*=\rho S_d/\rho_u$, heat release rate, and the stretch Karlovitz number $Ka_s$, all presented in nondimensional forms. 

As the primary cellular structure grows and widens, the local curvature at the cell center decreases, flattening the flame front \cite{berger2023flame}. This geometric flattening reduces the local stretch rate at the cell center below the values at the edges. Owing to the strong TD instability of lean hydrogen mixtures, the reductions in curvature and stretch rate at the flattened center weaken the local reaction intensity. Consequently, the flame displacement speed at the cell center decreases, the center to lag behind the edges where the reaction intensity is stronger. The advancing edges subsequently overtake the decelerating center, leading to a reversal of local convexity. The negative curvature at the cell center further suppresses local reactivity and reduces the displacement speed. This positive feedback between changes in displacement speed and flame-front curvature ultimately drives the splitting of a single cell into two smaller structures.

\begin{figure}[h!]
\centering
\vspace{10 pt}
\includegraphics[width=0.5\columnwidth]{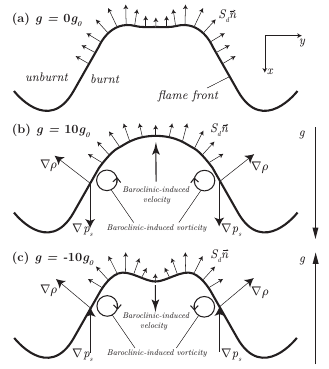}
\caption{Schematic of the baroclinic torque mechanism under different gravity conditions. The pressure-gradient vector  arising from the hydrostatic pressure force, denoted as \(\nabla p_s\), is related to the body-force field only ~\cite{sinibaldi1998suppression}.}
\label{fig:split_mechanism}
\end{figure}

The influence of gravity on cell splitting can be explained by the baroclinic torque mechanism, as illustrated in Fig.~\ref{fig:split_mechanism}. Baroclinic torques arise when the pressure gradient is not aligned with the density gradient ~\cite{drazin1981hydrodynamic, sinibaldi1998suppression}. Mathematically, they are represented by \( \nabla\rho \times \nabla p/\rho^2\), and physically they act as a vorticity source that generates additional momentum, alters the local curvature, and thus changes the angle between the pressure and density gradients. As shown in Fig.~\ref{fig:split_mechanism}(a), in the absence of gravity the central portion of the cell flattens before splitting, with concurrent decreases in the local stretch rate and displacement speed, causing the center to lag behind the edges. Under positive gravity (e.g., $g = 10 g_0$, as shown in Fig.~\ref{fig:split_mechanism}(b)), the pressure and density gradients form a blunt angle, generating a vortex pair whose induced velocities near the cell center point toward the unburned side of the flame. This accelerates local flame propagation, counteracts the convexity reversal at the cell center, and delays cell splitting, allowing larger cells to develop. In contrast, under negative gravity (e.g., $g = -10 g_0$, as shown in Fig.~\ref{fig:split_mechanism}(c)), the vortex-induced velocities point toward the burned side, promoting convexity reversal, cell splitting, and the formation of new, smaller cellular structures.

Note that the present study is not the only report on gravity effects on cell splitting. In a previous study by Sinibaldi et al. ~\cite{sinibaldi1998suppression}, it was observed that negative gravity can completely suppress flame wrinkles under certain conditions. However, their study examined propane and methane flames, which are TD-stable and near-neutral under lean conditions, respectively, with both DL and TD instabilities suppressed by negative gravity. By contrast, the present study focuses on lean hydrogen/air flames in which TD instability is strong. Consequently, even under negative gravity, the flame remains unstable; the influence of negative gravity manifests as accelerated cell splitting and delayed formation of large-scale cellular structures.

\subsubsection{Effects of gravity on the probability distributions of cell size, displacement speed, Karlovitz number and local curvature} \label{sec:cell_size}

\begin{figure}[h!]
\centering
\vspace{10 pt}
\includegraphics[width=0.75\columnwidth]{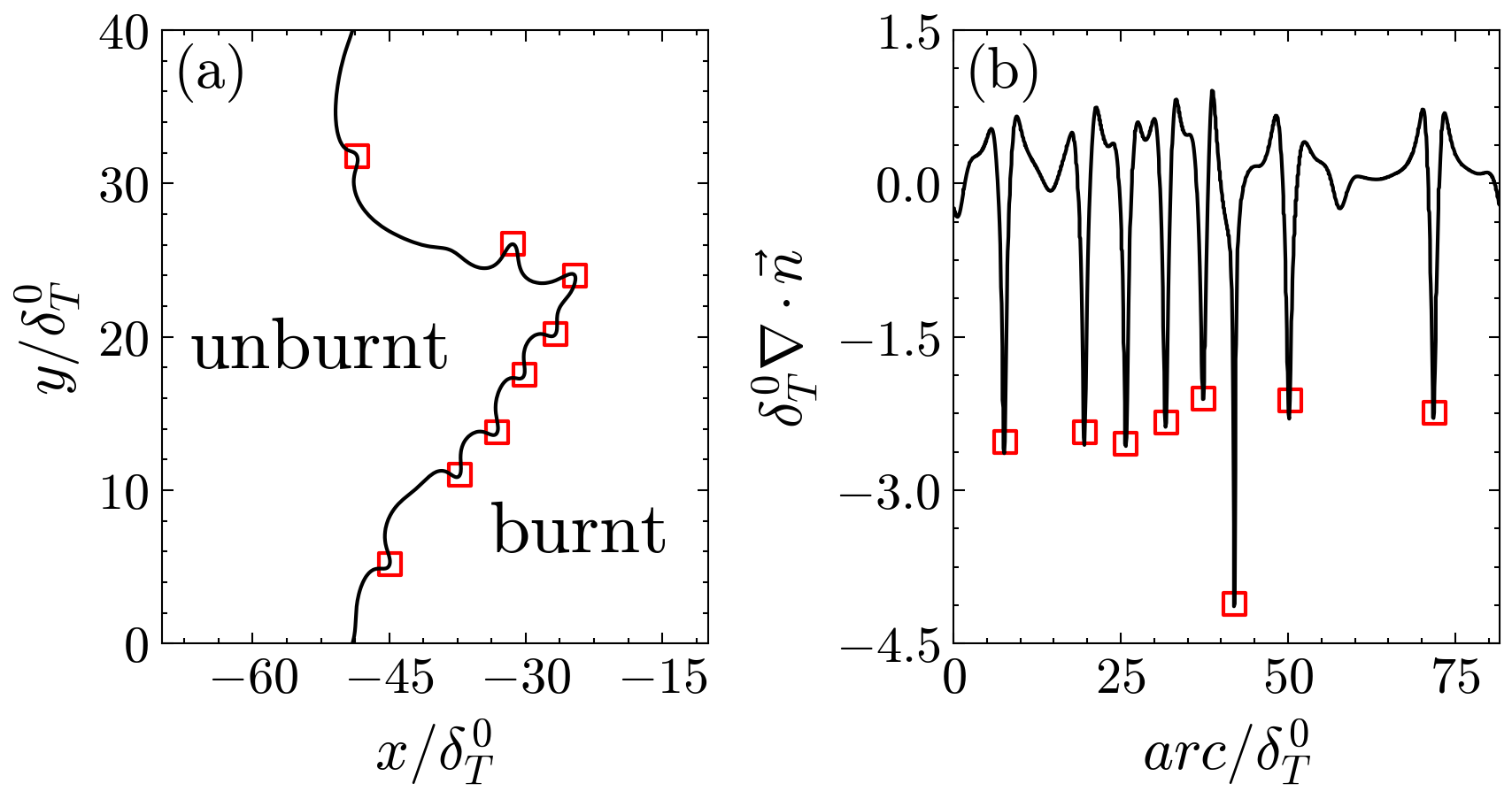}
\caption{Determination of the cell size $\lambda_{\text{cell}}$. (a) Morphology of the flame front, (b) curvature along the arc length of the flame front, with cusps marked by red squares.}
\label{fig:flamefront_and_curvature}
\end{figure}

In this section, we examine the influence of gravity on the statistical characteristics of cellular structures in fully developed flame instabilities within the nonlinear regime. The cell size \( \lambda_{\text{cell}} \) is defined as the distance between the tips of two consecutive cusps ~\cite{berger2019characteristic}. Fig.~\ref{fig:flamefront_and_curvature} shows the flame front morphology in the nonlinear regime and the dimensionless curvature along the arc length, with curvature cusps marked by red squares in both panels. Using a semicircular-arc approximation, the cell size is estimated as \( \lambda_{\text{cell}} = 2l^*/\pi \), where \( l^* \) denotes the arc length between two successive cusps along the flame front.

\begin{figure}[h!]
\centering
\vspace{10 pt}
\includegraphics[width=0.5\columnwidth]{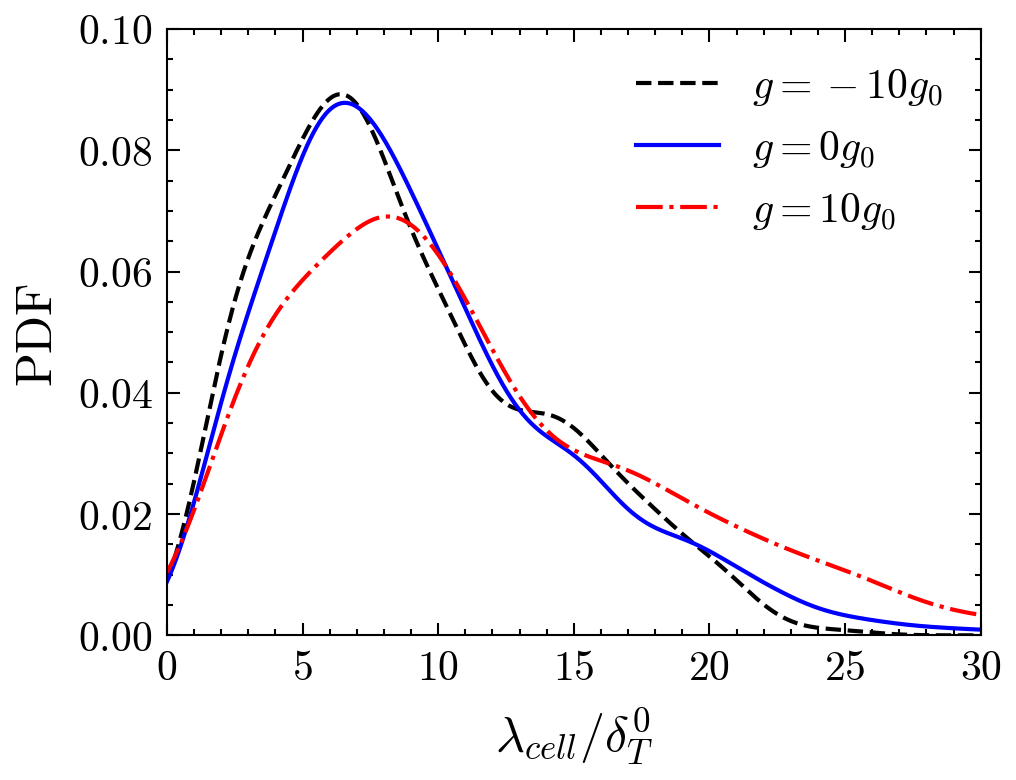}
\caption{PDF of the dimensionless cellular structure size, $\lambda_{\text{cell}}/\delta_T^0$, under various gravity conditions. The results indicate that positive gravity (RT-unstable) increases the most probable cell size and shifts the PDF toward larger values.} \label{fig:Cell_size_PDF}
\end{figure}

Fig.~\ref{fig:Cell_size_PDF} shows the Probability Density Function (PDF) of \( \lambda_{\text{cell}} \) under different gravity conditions. In the absence of gravity, the most probable cell size is \( \lambda^{mp}_{\text{cell}} \approx 6 \delta_T^0 \), corresponding to a dimensionless wavenumber \( \bar k^{mp} = 1.08 \). This value is very close to the observed peak wavenumber \( \bar k_{\text{max}} = 1.10 \) from the linear dispersion relation shown in Fig.~\ref{fig:six_fig}(b). The result is consistent with prior studies on lean hydrogen/air flames and their cellular structures ~\cite{berger2019characteristic, yan2026quantitative}.

Under positive gravity, the PDF of cell size shifts to the right side, with the most probable cell size occurring around \(8 \delta_T^0 \). The probability of large cells (\( \lambda_{\text{cell}} > 15 \delta_T^0 \)) is also significantly higher, because positive gravity enhances cell stability and increases the critical size required for cell splitting. Under negative gravity, the PDF of cell size is essentially indistinguishable from the zero-gravity case.

\begin{figure}[h!]
\centering
\vspace{10 pt}
\includegraphics[width=0.75\columnwidth]{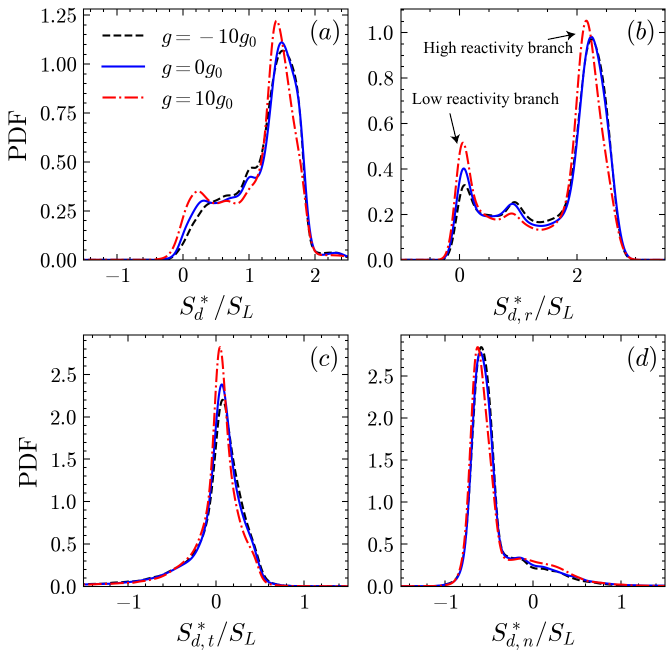}
\caption{PDFs of the density-weighted displacement speed $S_d^*$ and its three constituent components: the reaction term $S_{d,r}^*$, the normal diffusion term $S_{d,n}^*$, and the tangential diffusion term $S_{d,t}^*$.}
\label{fig:Sd4pdf}
\end{figure}

Fig.~\ref{fig:Sd4pdf} presents the probability density functions (PDFs) of the density-weighted displacement speed $S_d^*$ and its three components, $S_{d,r}^* = \rho S_{d,r}/\rho_u$, $S_{d,n}^* = \rho S_{d,n}/\rho_u$, and $S_{d,t}^* = \rho S_{d,t}/\rho_u$. Due to the strong TD instability, the peak value of $S_d^*$ deviates from $S_L$ under all gravity conditions. For the reaction component $S_{d,r}^*$, the PDF exhibits two branches corresponding to higher- and lower-reactivity regions, which align with the convex and flat/concave portions of the cellular structure, respectively. The $g=10g_0$ case shows a higher probability density on the low-reactivity branch, which indicates enhanced stability near the cell-splitting threshold and is consistent with the prior analysis. The peak probability density of the tangential diffusion component $S_{d,t}^*$ shows a positive correlation with gravity, while the PDF of the normal diffusion component $S_{d,n}^*$ remains nearly the same across the three gravity conditions.

\begin{figure}[h!]
\centering
\vspace{10 pt}
\includegraphics[width=0.75\columnwidth]{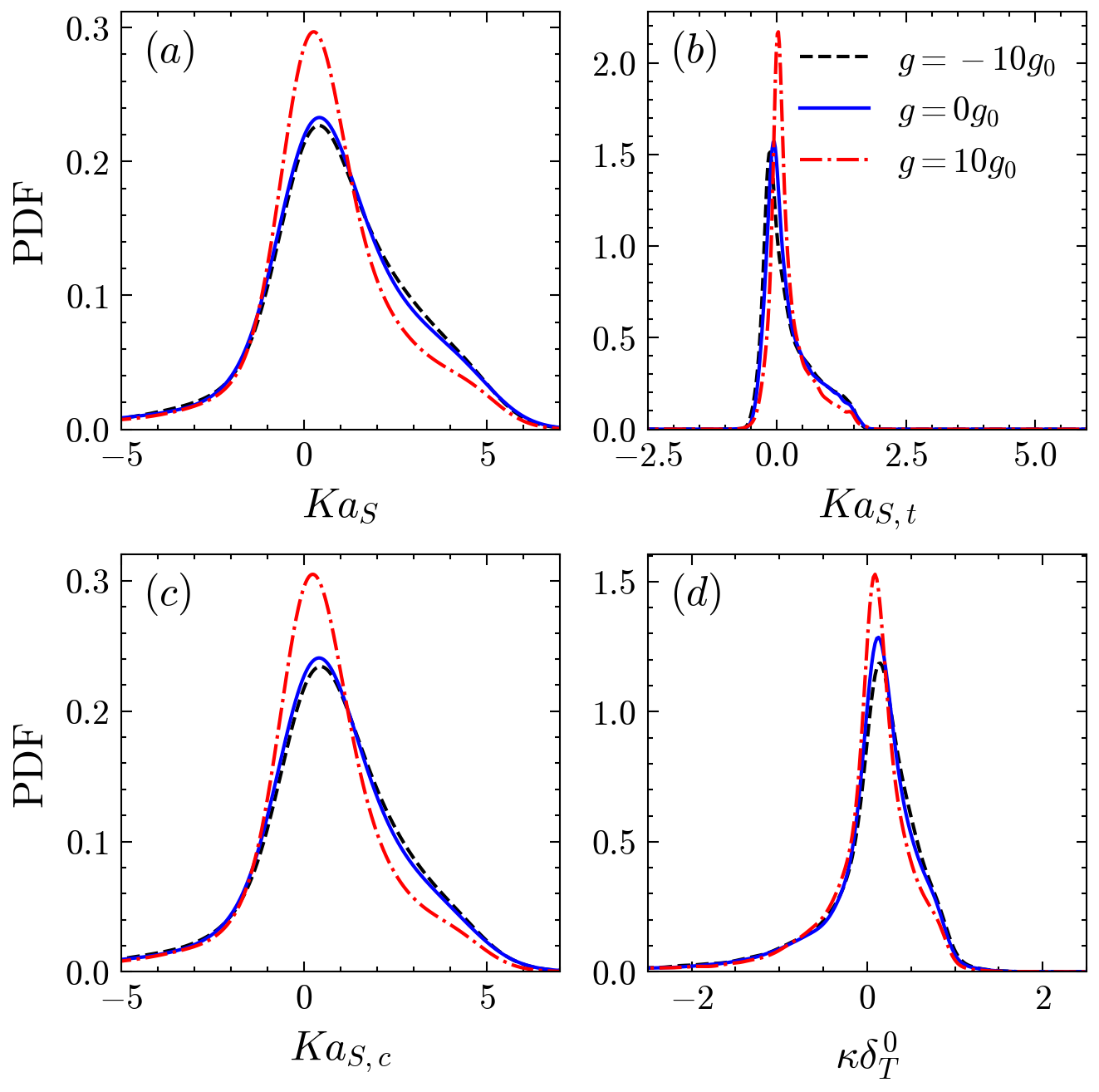}
\caption{PDFs of the stretch Karlovitz number $Ka_s$ and its decomposed components $Ka_{s,t}$ and $Ka_{s,c}$, together with the nondimensional local curvature $\kappa \delta_T^0$.}
\label{fig:K4pdf}
\end{figure}

Fig.~\ref{fig:K4pdf} shows the PDF of the Karlovitz number $Ka_s$, together with its two components $Ka_{S,t}$ and $Ka_{S,c}$, and the nondimensional curvature $\kappa \delta_T^0$. At \(g=10g_0\), the PDFs of all components near zero-stretch are notably higher than in the zero-gravity case. The variances of the PDFs decrease with gravity, suggesting that small-scale wrinkles are suppressed and the flame surface becomes smoother at elevated gravity levels. 

\subsubsection{Effects of gravity on the characteristic size of finger-like structures} \label{sec:flame_speed}

As individual cellular structures undergo continuous splitting and merging, the flame also develops finger-like structures on a larger scale. In the present study, gravity is found to influence the morphology of these finger-like features as well.

\begin{figure}[h]
\centering
\vspace{10 pt}
\includegraphics[width=0.5\columnwidth]{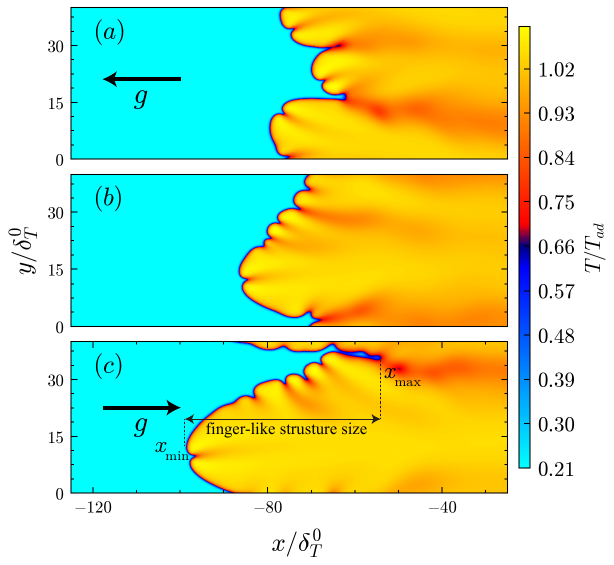}
\caption{Flame morphology at a fully developed state ($t/\tau_f=50$) for $\phi=0.4$ under gravity levels of (a) $g=-10g_0$ (RT-stable), (b) $g=0$, and (c) $g=10g_0$ (RT-unstable).}
\label{fig:nonlinear_regime_1}
\end{figure}

Fig. \ref{fig:nonlinear_regime_1} shows the fully developed flame morphology under different gravity conditions. The local temperature, normalized by the adiabatic flame temperature of a planar flame, is plotted to aid visualization of the flame front. Because the lean hydrogen/air mixture has an effective Lewis number less than unity ($Le = 0.34$), the deficient reactant preferentially diffuses toward convex flame segments (the flame tips) facing the unburned gas. This local enrichment enhances reaction rates, yielding super-adiabatic temperatures at the flame tips. More importantly, increasing the effective gravity level promotes the formation of larger finger-like structures.

\begin{figure}[h]
\centering
\vspace{10 pt}
\includegraphics[width=0.5\columnwidth]{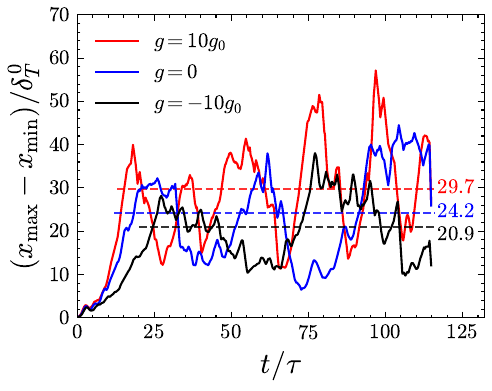}
\caption{Temporal evolution and time-averaged value of the characteristic finger size observed in a flame at $\phi=0.4$, under different gravity conditions. RT instability promotes the formation of larger fingers.}
\label{fig:finger_size}
\end{figure}

Quantitatively, the characteristic size of finger-like structures is defined as the distance between the leading and trailing edges of the flame front, as shown in Fig.~\ref{fig:nonlinear_regime_1}(c). The temporal evolution of finger size under various gravity conditions is illustrated in Fig. ~\ref{fig:finger_size}. While the instantaneous finger size exhibits strong temporal fluctuations due to intense flame instabilities, the time-averaged value (marked by the dashed lines) clearly indicates that positive gravity markedly promotes the growth of finger-like structures, whereas negative gravity suppresses their growth.

\subsubsection{Effects of gravity on the flame consumption speed}

The present study also finds that positive gravity can also increase the the flame consumption speed \(S_c\). \(S_c\) is a global quantity defined by the integrated consumption rate of \ce{H2} over the entire computation domain \(\Omega\), i.e., 
\begin{equation}
    S_c = -\frac{1}{\rho_u Y_{\ce{H2},u} L_y} \iint_\Omega \dot{\omega}_{\ce{H2}} \, \mathrm{d}x \, \mathrm{d}y.
\end{equation}
Here, \(\rho_u\) denotes the density of the unburned gas, \(Y_{\ce{H2},u}\) is the mass fraction of \ce{H2} in the unburned gas, and \(\dot{\omega}_{\ce{H2}}\) is the fuel consumption rate.  
According to Berger et al. \cite{berger2023flame}, changes in the flame consumption rate are primarily governed by two mechanisms: (1) an increase in the surface area \(A/L_y\) due to flame wrinkling, and (2) variations in reaction intensity induced by flame stretching under preferential diffusion, which is represented by the stretch factor \(I\). The overall effect can be summarized as follows:
\begin{equation}
    \frac{S_c}{S_L} = \frac{A}{L_y} I
\end{equation}

In the present study, the flame surface area $A$ is determined using the generalized flame surface density (GFSD) formalism \cite{vervisch1995surface}, based on the progress variable $c$:
\begin{equation}
A = \iint_\Omega \left| \nabla c \right| \ \mathrm{d}x  \mathrm{d}y
\end{equation}

\begin{figure}[h]
\centering
\vspace{10 pt}
\includegraphics[width=0.5\columnwidth]{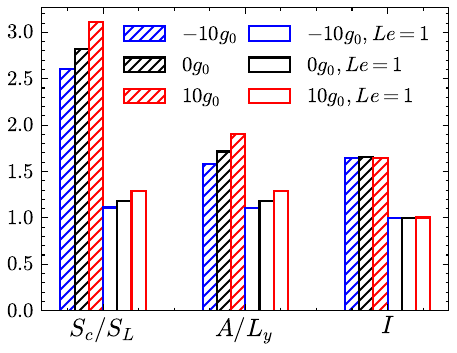}
\caption{Normalized flame speed $S_c/S_L$, flame front area $A/L_y$, and stretch factor $I$ for different gravity levels at \(\phi=0.4\). Results are shown for both $Le = 0.34$ (striped bars) and $Le = 1$ (white bars).}
\label{fig:Sc_bar}
\end{figure}

Figure~\ref{fig:Sc_bar} shows the time-averaged \( S_c/S_L \), \( A/L_y \), and \( I\) under different gravity conditions. For all cases, positive (negative) gravity significantly increases (decreases) the flame consumption speed \( S_c \). The change in $S_c$ is mostly caused by the flame area \( A \), as the stretch factor \( I \) remains nearly constant across all gravity conditions. This observation is inherently consistent with the effect of gravity on finger-like structures. Specifically, the development of larger fingers under positive gravity substantially increases the flame surface area.

Note that gravity can affect $S_c$ even in the absence of TD instability. For reference, we have conducted additional simulations in which a unity Lewis number is artificially set to eliminate the effects of preferential diffusion. A similar trend is observed at $Le = 1$, that gravity affects $S_c$ mainly through the flame area $A$, whereas the stretch factor $I$ has little dependence on gravity.

\section{Conclusions} \addvspace{10pt}
Time-resolved simulations incorporating detailed reaction chemistry and transport are conducted to investigate the instability of hydrogen/air flames in both linear and nonlinear regimes under various gravity conditions. In the linear regime, the influence of gravity on the growth rate of disturbance is found to be most significant under ultra-lean, low-temperature and high-pressure conditions. A global gravity-sensitivity parameter, $\eta$, is introduced to quantify the influence of gravity on the maximum instability growth rate, and a universal scaling law is established between $\eta$ and the Froude number $Fr$ across various equivalence ratios, temperatures and pressures: $\eta \sim Fr^{-0.69}$. The influence of gravity on flame instability is negligible at $Fr > 10$. 

In the nonlinear regime, gravity is found to have opposite effects on the large-scale and small-scale structures of the flame front. For the small-scale cellular structures, positive gravity inhibits cell splitting via a baroclinic torque mechanism, thereby promoting the growth of smoother, larger cells and reducing the average stretch rate of the flame. Quantitative analyses of the PDFs of cell size, displacement speed, Karlovitz number, and local curvature yield results consistent with this observation. For the large-scale finger-like structures, positive gravity increases the finger size, thereby increasing the total surface area and the global consumption speed of the flame. The stretch factor, however, is insensitive to gravity. The findings of the present study should be useful for both the fundamental understanding of hydrogen flame dynamics and relevant applications such as fire safety and space propulsion.

\begin{acknowledgments}
This work was supported by the Space Application System of China Manned Space Program, by the National Natural Science Foundation of China under Grant No. 52425604, and by the National Key Research and Development Program of China under Grant No. 2025YFF0511801.
\end{acknowledgments}

\section*{Data availability statement}
The data that support the findings of this article are openly available \cite{Wen2026}.

\bibliography{apssamp}

\end{document}